\newcommand\papertitle{Optimal bispectrum estimator and simulations of the the CMB Lensing-ISW non-Gaussian signal}

\documentclass[twocolumn,traditabstract]{aa} 
\usepackage{graphicx}
\usepackage{txfonts}
\usepackage{url}
\usepackage{natbib}
\bibpunct{(}{)}{;}{a}{}{,}  
\usepackage{color}
\usepackage{colortbl} 

 \def\be{\begin{equation}}
 \def\ee{\end{equation}}
 \def\bea{\begin{eqnarray}}
 \def\eea{\end{eqnarray}}

\begin{document}
%

\title{\papertitle}

\offprints{mangilli@iap.fr}
\authorrunning{Mangilli, Wandelt, Elsner \& Liguori}
\titlerunning{Optimal estimator and simulations of the the CMB Lensing-ISW bispectrum}


\author{A. Mangilli \inst{\ref{inst1}} 
\and B. Wandelt  \inst{\ref{inst1},\ref{inst1a} } 
\and Franz Elsner   \inst{\ref{inst1}} 
\and Michele Liguori  \inst{\ref{inst2}} } 

\institute{Institut d'Astrophysique de Paris et Universit\'e Pierre et Marie Curie Paris 6, 
98bis Bd. Arago 75014 Paris,
France \\
\email{mangilli@iap.fr} \label{inst1}
\and
International Chair of Theoretical Cosmology,
Lagrange Institute (ILP)
98 bis, boulevard Arago
75014 Paris
France\label{inst1a}
\and
INFN, Sezione di Padova and Dipartimento di Fisica e Astronomia ÒG. GalileiÓ,
Universit\'a degli Studi di Padova, Via Marzolo 8, 35131 Padova, Italy \label{inst2}}


\abstract{In this paper we present the tools to optimally extract the Lensing-Integrated Sachs Wolfe (L-ISW) bispectrum signal from future CMB data. 
We implement two different methods to simulate the non-Gaussian CMB maps with the L-ISW signal: a non-perturbative method based on the FLINTS lensing code and the separable mode expansion method.
We implement the Komatsu, Spergel and Wandelt 
(KSW) optimal estimator analysis for the Lensing-ISW bispectrum and we test it on the non-Gaussian simulations in the case of a realistic CMB experimental settings with an inhomogeneous sky coverage. 
We show that the estimator approaches the Cramer-Rao bound and that Wiener filtering the L-ISW simulations gives a slight improvement on the estimate of $f_{NL}^{L-ISW}$ of $\leq 10\%$. 
For a realistic CMB experimental setting accounting for anisotropic noise and masked sky, we show that the linear term of the estimator is highly correlated to the cubic term 
and it is necessary to recover the signal and the optimal error bars. 
We also show that the L-ISW bispectrum, if not correctly accounted for, yields  an underestimation of the $f_{NL}^{local}$ error bars of $\simeq 4\%$. 
 A joint analysis of the non-Gaussian shapes and/or L-ISW template subtraction is needed in order to recover unbiased results of the primordial non-Gaussian signal from ongoing and future CMB experiments.
}


\maketitle
%


\section{Introduction}

One of the most relevant mechanisms that can generate non-Gaussianity from secondary Cosmic Microwave Background (CMB) anisotropies is the coupling between weak lensing and the Integrated Sachs Wolfe (ISW) \citep{1967ApJ...147...73S} 
 and the Rees Sciama (RS) \citep{1968Natur.217..511R}.
 This correlation gives in fact the leading contribution to the CMB secondary bispectrum with a blackbody frequency dependence \citep{Goldberg:1999xm,Verde:2002mu,Giovi:2004te}. 
Weak lensing of the CMB is caused by gradients in the matter gravitational potential that distorts the CMB photon geodesics. The ISW and the RS effects, on the other hand, are related to the time variation of the gravitational potential wells. 
  The relevant mechanism is given by the late ISW, owing to the action of Dark Energy which causes the decay of the gravitational potential wells  as the Universe expands.  
Both the lensing and the ISW effect are then related  
to the matter gravitational potential and thus are correlated phenomena. 
This gives rise to a non-vanishing three-point correlation function or, analogously, a non-vanishing bispectrum, its Fourier counterpart.
 The RS (also referred as the non-linear ISW) arises when the growth of structure in the evolving universe becomes non-linear. Being a second order effect, the RS gives a smaller contribution to the signal with respect to the ISW.
The CMB bispectrum arising from the cross correlation between lensing and ISW/RS (from now on referred to as L-ISW) is expected to have an high signal-to-noise ratio from ongoing and future CMB experiments so that it will be detectable in the near future with an high statistical significance \citep{Verde:2002mu,Giovi:2004te,Mangilli:2009dr,Lewis:2011fk}. 
A detection would open the possibility to exploit the cosmological information related to the late time evolution encoded in the L-ISW signal. 
It is useful to stress that a significant detection of the L-ISW signal from ongoing CMB experiments like Planck would be a powerful probe of Dark Energy from CMB alone and it would be a complementary probe of the late time Universe with respect to the large scale structure and the the CMB power spectrum analysis. 
 Moreover,  \citet{Mangilli:2009dr,Hanson:2009kg} showed that the L-ISW bispectrum can be a serious contaminant problem for the estimation of the primary local non-Gaussianity from future data.
Ongoing CMB experiment such as Planck \citep{Ade:2011ah} and future experiments like COrE \citep{Bouchet:2011ck} will then require a detailed reconstruction of the L-ISW bispectrum
 either to be able to correctly remove the L-ISW contribution when estimating the local primary non-Gaussian parameter $f_{NL}$, or to exploit the cosmological information encoded in the signal; 
therefore it becomes extremely important to know how to model and simulate it. 

In this paper we present the formalism and the numerical implementation {\it i}) to generate simulated CMB maps containing the L-ISW signal 
and {\it ii}) to build and test the optimal estimator for the L-ISW bispectrum, accounting for both the cubic and the linear parts. 
The linear part for this specific kind of signal, has been here calculated and tested for the first time.
As regarding the CMB non-Gaussian simulations, we implemented and tested the L-ISW signal with two methods: the separable mode expansion method \citep{Fergusson:2009nv,Smith:2006ud} %
and the non-perturbative approach described in Sec. \ref{sec:COV}.

It is important to have an optimal estimator for the L-ISW bispectrum in order to extract the signal optimally from future data and to disentangle it from other kinds of non-Gaussianities, i.e. the local primary bispectrum, with which it is degenerated. Here, following \citet{Komatsu:2010hc} and \citet{Munshi:2009fr}, we implemented the KSW bispectrum estimator \citep{Komatsu:2003iq} for the L-ISW signal of a full sky, cosmic variance limited CMB experiment and in the case of a more realistic instrumental setting, similar to that of a space-based experiment. 
Furthermore, for this realistic case, we investigate the statistical detection significance and the impact that the L-ISW bispectrum has on the estimation and on the variance of the primary local non-Gaussian parameter $f_{NL}$. %

The outline of this paper is as follows. 
In section \ref{sec:sims} we present the methods to simulate the non-Gaussian CMB maps containing the L-ISW bispectrum signal by the use of both the separable mode expansion method and the non-perturbative covariance method. 
Section \ref{sec:estISW} provides the basics to build and implement the optimal estimator for the L-ISW signal, including its linear part.  It also includes a discussion regarding the implementation of the Wiener filtered simulations algorithm. %
In section \ref{sec:results} we present the relevant tests and results. 
In section \ref{sec:err} we quantify the statistical detection significance of the L-ISW bispectrum and the impact on the error of primary non-Gaussianity $f_{NL}$ due to the presence of the ISW signal. Finally, in section \ref{sec:conclusions}, we discuss the results and we summarize the conclusions. Details on the simulations built with the covariance method and on the L-ISW cross correlation coefficients are given in the appendix.

\section{Simulated non-Gaussian CMB maps}\label{sec:sims} 

In this section, we present the formalism to create simulated CMB maps for the L-ISW bispectrum.  We use two different methods:  a no perturbative approach, here named the 'covariance method',  
and the separable modes expansion method (\citep{Fergusson:2009nv} and \citep{Smith:2006ud}). The latter gives an efficient and easy to handle way to generate L-ISW maps, while the former method provides better insights on the physics related to the L-ISW bispectrum. In this case, in fact, the L-ISW signal is generated starting from the covariance matrix representing the expected correlation between the lensing and the ISW/RS effects.

\subsection{Covariance method}\label{sec:COV}

The L-ISW correlation is defined by the covariance matrix:
\be
\mathbf{C}_{L-ISW}= \left(
  \begin{array}{cc}
  C_\ell^{\phi \phi}& C^{T \phi}_\ell \\               
  C^{T \phi}_\ell&C_\ell^{TT}
  \end{array}
  \right)
\ee
and the cross correlation coefficient:
\be
\label{eq:r}
r^{T\phi}=\frac{ C^{T \phi}_\ell}{ \sqrt{C_\ell^{TT}}   \sqrt{C_\ell^{\phi \phi}}  }.
\ee
Here, $C_\ell^{TT}  \delta_{\ell \ell'} \delta_{m m'}= \langle   a^P_{\ell m}  a^{P*}_{\ell' m'}\rangle$ and $C_\ell^{\phi \phi} \delta_{\ell \ell'} \delta_{m m'} =  \langle   \phi^L_{\ell m}  \phi^{L*}_{\ell' m'}\rangle$ are respectively the CMB primary temperature power spectrum and the lensing power spectrum, where the lensing potential $\phi$ (the gravitational potential projection along the line of sight) is defined by:
\be
\phi(\hat{\mathbf{n}})= -2 \int^{r_{ls}}_0 dr \frac{r(z_{ls})-r(z)}{r(z) \, r(z_{ls}) } \, \Phi(r, \hat n r).
\ee
The term in the numerator, $ C^{T \phi}_\ell$, is the power spectrum of the cross correlation between the lensing and the ISW/RS effect, see appendix \ref{app:Cltphi} for details. %

After a Cholesky decomposition of the L-ISW correlation matrix $\mathbf{C}_{L-ISW}$, the two new variables $t_{\ell m}$ and $z_{\ell m}$ are then defined by
\bea
t_{\ell m}&=&\sqrt{C_\ell^{\phi \phi}}x_{\ell m} \equiv  \phi^L_{\ell m} \\
z_{l m}&=&\sqrt{C_{\ell}^{TT}}[x_{\ell m} r^{T\phi}+ y_{\ell m}\sqrt{1-(r^{T\phi})^2}],
\label{eq:zlm}
\eea
where $x_{\ell m}$ and $y_{\ell m}$ are two independent random gaussian fields. By definition, the new fields are such that:
$\langle t^2 \rangle= C_\ell^{\phi \phi}$, $\langle z^2 \rangle= C_\ell^{TT}$ and they have the non-zero cross correlation  $\langle z t \rangle=  C^{T \phi}_\ell$.

\begin{figure}[b!]
\includegraphics[width=8.8cm,height=6.5cm]{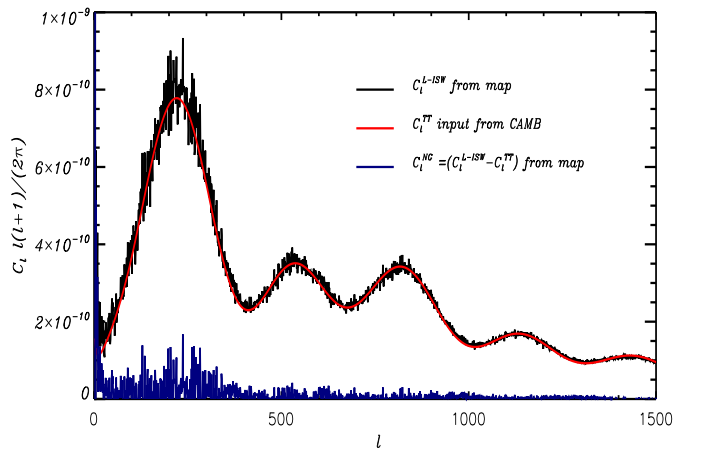} 
\caption{{\it The L-ISW power spectrum from the covariance method simulation.} The plot shows that the temperature power spectrum of the L-ISW simulations generated with the method described in Sec.~\ref{sec:COV} is compatible with the input theoretical power spectrum from CAMB and that the non-Gaussian contribution is always subdominant. The temperature power spectrum from one simulated L-ISW realization is shown in black, the red line refers to the theoretical input from CAMB while the blue refers to the non-Gaussian L-ISW contribution from the same realization.}
\label{fig:ClsCOV}
\end{figure} 

As described in appendix  \ref{app:ISWcov}, the map that contains the desired L-ISW bispectrum is then given by the coefficients
 \be
 a^{L-ISW}_{\ell m}=z_{\ell m} + a^L_{\ell m} - a^P_{\ell m} \equiv  z_{\ell m} + \Delta a^L_{\ell m},
 \label{eq:ISWalm}
 \ee
 where $ a^P_{\ell m}$ and  $a^L_{\ell m}$ are, respectively, the unlensed primary and the lensing angular coefficients and $\Delta a^L_{\ell m} =a^L_{\ell m} - a^P_{\ell m}$
 corresponds to the lensing expansion terms only.
Note that by construction $ y_{\ell m}$ has the same phases as $a^P_{\ell m}\equiv  y_{\ell m}\sqrt{ C_{\ell}^{TT}}$ and $\phi^L_{\ell m}\equiv  x_{\ell m}\sqrt{ C_{\ell}^{\phi \phi}}$ the same as $x_{\ell m}$, which is necessary for building a map with the wanted bispectrum signal.

Figure \ref{fig:ClsCOV} shows in black the temperature CMB power spectrum 
of one simulated L-ISW map, $C_\ell^{L-ISW}$, built from Eq. \ref{eq:ISWalm}. The non-Gaussian contribution, in blue in the figure, is always subdominant and the $C_\ell^{L-ISW}$ are consistent with the theoretical input $(C_\ell^{TT})_{th}$ (red line) obtained with CAMB \citep{Lewis:1999bs} \footnote{http://camb.info}. 
As throughout the paper, the reference cosmological model used is the $\Lambda$CDM model with parameter values defined in \citep{Komatsu:2010fb}.

\begin{figure}[t!]
\includegraphics[width=8.8cm,height=6.5cm]{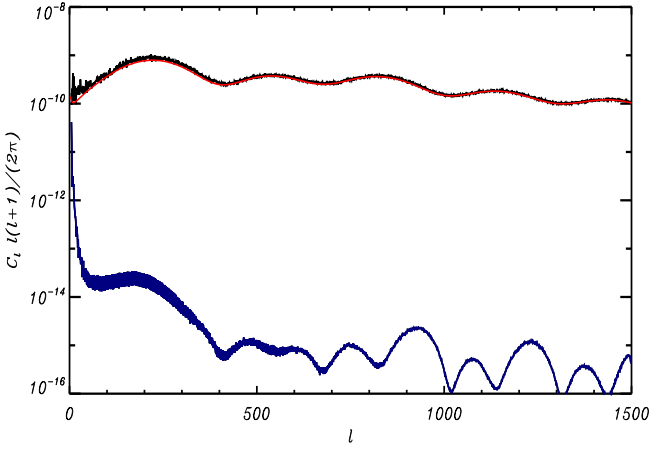} 
\caption{ 
{\it The L-ISW power spectrum from the separable mode expansion method simulation.} The temperature power spectrum of the L-ISW simulation generated with the method described in Sec.~\ref{sec:NGmethod} is compatible with the input theoretical power spectrum from CAMB and  the non-Gaussian contribution is always subdominant. The temperature power spectrum from one simulated L-ISW realization is shown in black, the red line refers to the theoretical input from CAMB while the blue refers to the non-Gaussian L-ISW contribution from the same realization.
}
\label{fig:clsNG}
\end{figure}

\subsection{Separable modes expansion method}\label{sec:NGmethod}
Following  \citet{Fergusson:2009nv} and \citet{Smith:2006ud}, 
the non-Gaussian part of the CMB angular coefficients can be defined starting from a given reduced bispectrum.
In the case of the L-ISW signal, the method can be used because this kind of signal is separable, so
\bea
[a^{NG}_{\ell m}]_{L-ISW}= \int d^2 \hat{\bf n} \sum_{\ell_2 m_2 \ell_3 m_3}&& b_{\ell_1 \ell_2 \ell_3}^{L-ISW} Y_\ell^m(\hat{\bf n}) \nonumber\\
&& \frac{ a^G_{\ell_2 m_2} Y_{\ell_2}^{m_2}(\hat{\bf n})}{C_{\ell_2}}  
\frac{ a^G_{\ell_3 m_3} Y_{\ell_3}^{m_3}(\hat{\bf n})}{C_{\ell_3}}.
\eea

From the expression of the L-ISW reduced bispectrum in Eq. (\ref{eq:ISW}) and by factorizing the $\ell$ dependence, the explicit form of the non-Gaussian contribution to the $a_{\ell m}$ from the L-ISW cross correlation is given by:
\bea
\label{eq:almNG}
[a^{NG}_{\ell m}]_{L-ISW} &=& \frac{1}{6} \int d^2 \hat{\bf n} Y_\ell^m(\hat{\bf n}) \Big[ \ell(\ell+1) Q( \hat{\bf n}) E( \hat{\bf n})  \nonumber\\
&+&C_\ell \big( [\delta^2E]( \hat{\bf n}) Q( \hat{\bf n}) - [\delta^2Q]( \hat{\bf n}) E( \hat{\bf n}) \big) \nonumber \\
&-& \big( [\delta^2P]( \hat{\bf n}) Q( \hat{\bf n}) 
+  [\delta^2Q]( \hat{\bf n}) P( \hat{\bf n}) \big) \nonumber\\
&-& \ell(\ell+1) Q( \hat{\bf n}) P( \hat{\bf n}) \\
&+& q_\ell \big( [\delta^2E]( \hat{\bf n}) P( \hat{\bf n}) - [\delta^2P]( \hat{\bf n}) E( \hat{\bf n}) \big) \nonumber\\
&+&  \ell(\ell+1) q_\ell P( \hat{\bf n}) E( \hat{\bf n})
 \Big]. \nonumber
\eea
Here,
\bea
&&P(\hat{\bf{n}}) \equiv\sum_{\ell m}  a_{\ell m}Y_{\ell m}(\hat{\bf{n}}), \nonumber\\
&&Q(\hat{\bf{n}})\equiv \sum_{\ell m} C^{T \phi}_\ell  (C^{-1} a)_{\ell m} Y_{\ell m}(\hat{\bf{n}}),  \\
&&E(\hat{\bf{n}})\equiv\sum_{\ell m} (C^{-1} a)_{\ell m} Y_{\ell m}(\hat{\bf{n}}) \nonumber
\label{eq:maps}
\eea
The maps with a $\delta^2$ prefix are given by, e.g.,
$\delta^2 P = -\sum_{\ell}\ell(\ell+1) a_{\ell m}Y_{\ell m}(\hat{\bf{n}}) $; they correspond to the maps of Eq. \ref{eq:maps} multiplied by the $-\ell(\ell+1)$ factor.
The final solution containing the L-ISW signal is then:
\be
a_{\ell m}= a^G_{\ell m} + [a^{NG}_{\ell m}]_{L-ISW},
\ee
where $ a^G_{\ell m}$ is the Gaussian part.

In Fig. \ref{fig:clsNG}, we show the CMB temperature power spectra from the Gaussian and the non-Gaussian map, as defined in Eq. (\ref{eq:almNG}). The non-Gaussian contribution is always subdominant as expected.

%
\section{The Optimal KSW estimator for the lensing-ISW/RS bispectrum}\label{sec:estISW}

In this section we present the formalism related to the KSW estimator \citep{Komatsu:2003iq} for the Lensing-Integrated-Sachs Wolfe bispectrum signal.

\subsection{Definition}

The $a_{\ell m}$ probability distribution function (PDF) in the limit of weak non-Gaussianity (i.e. truncated at the bispectrum level) is given by \citep{Babich:2005en,Taylor:2000hq,Komatsu:2010hc}:
\begin{eqnarray}
\nonumber
& &P(a)=
\frac1{(2\pi)^{N_{\rm harm}/2}|C|^{1/2}}
\exp\left[-\frac12\sum_{lm}\sum_{l'm'}a_{lm}^*(C^{-1})_{lm,l'm'}a_{l'm'}\right]\\  
\nonumber
&\times&
\left\{
1+\frac16\sum_{{\rm all}~l_im_j}
\langle a_{l_1m_1}a_{l_2m_2}a_{l_3m_3}\rangle
\left[
(C^{-1}a)_{l_1m_1}(C^{-1}a)_{l_2m_2}(C^{-1}a)_{l_3m_3}
\right.\right.\\
& &
\left.\left.
-3(C^{-1})_{l_1m_1,l_2m_2}(C^{-1}a)_{l_3m_3}
\right]\right\}.
\label{eq:pdf}
\end{eqnarray}
where $\langle a_{\ell_1m_1}a_{\ell_2m_2}a_{\ell_3m_3}\rangle$ is the {\it angular bispectrum}.
 Here, we are interested in the L-ISW case, for which the angular bispectrum, parametrized by the amplitude parameter $f_{NL}^{L-ISW}$, is
\begin{equation}
\langle a_{\ell_1m_1}a_{\ell_2m_2}a_{\ell_3m_3}\rangle={\cal G}_{\ell_1\ell_2\ell_3}^{m_1m_2m_3}       
f_{   NL      }^{L-ISW }
b_{\ell_1\ell_2\ell_3}^{L-ISW},
\end{equation}
where 
\be
 b_{\ell_1\ell_2\ell_3}^{L-ISW}
=
\left[
\frac{\ell_1(\ell_1+1)-\ell_2(\ell_2+1)+\ell_3(\ell_3+1)}2C_{\ell_1}^P
 C^{T \phi}_{\ell_3}
+(5p)\right],  %
\label{eq:ISW}
\ee
is the reduced bispectrum and $ C^{T \phi}_\ell\equiv\langle\phi^{*}_{\ell m}a_{\ell m}^{L-ISW}\rangle$ are the L-ISW
cross-correlation coefficients. %
According to \citep{Komatsu:2003iq}, for small departure from Gaussianity, the optimal estimator for theL-ISW amplitude parameter is given by:
\begin{equation}
 f_{NL}^{L-ISW} = (F^{-1})S_{L-ISW},
\label{eq:est}
\end{equation}
 where $(F^{-1})$ is the inverse of the L-ISW Fisher matrix
\begin{equation}
F \equiv F^{L-ISW}= f_{sky}\sum_{2 \leqslant \ell_1 \leqslant \ell_2 \leqslant \ell_3} \frac{ B^{L-ISW}_{\ell_1 \ell_2 \ell_3} \, B^{L-ISW}_{\ell_1 \ell_2 \ell_3} }{ \Delta_{\ell_1 \ell_2 \ell_3} C_{\ell_1}  C_{\ell_2}   C_{\ell_3} }.
\label{eq:FisherLISW}
\end{equation}
In the case of a realistic CMB experimental setting, the noise, $N_\ell$, and the beam window function, $w_\ell$, are accounted for so that $C_\ell=N_\ell+C_\ell^{th}w^2_\ell$. In this case, the bispectrum is also convolved with the beam transfer function $w_\ell$, $B^{L-ISW}_{\ell_1 \ell_2 \ell_3} \propto b^{L-ISW}_{\ell_1 \ell_2 \ell_3}w_{\ell_1}w_{\ell_2}w_{\ell_3} $. Given a mask $M( p )$, the observed sky fraction $f_{sky}$ is defined as: 
\be
f_{sky}=\frac{\sum_p M( p) }{N_{pix}},
\ee
where $N_{pix}=12 N_s^2$ is the number of pixels in the map, $N_s$ is the map resolution and the sum  $\sum_p$ is done over the pixels.

\begin{figure}[t!]
\centering
{\includegraphics[width=6.5cm,height=8.8cm,angle=90]{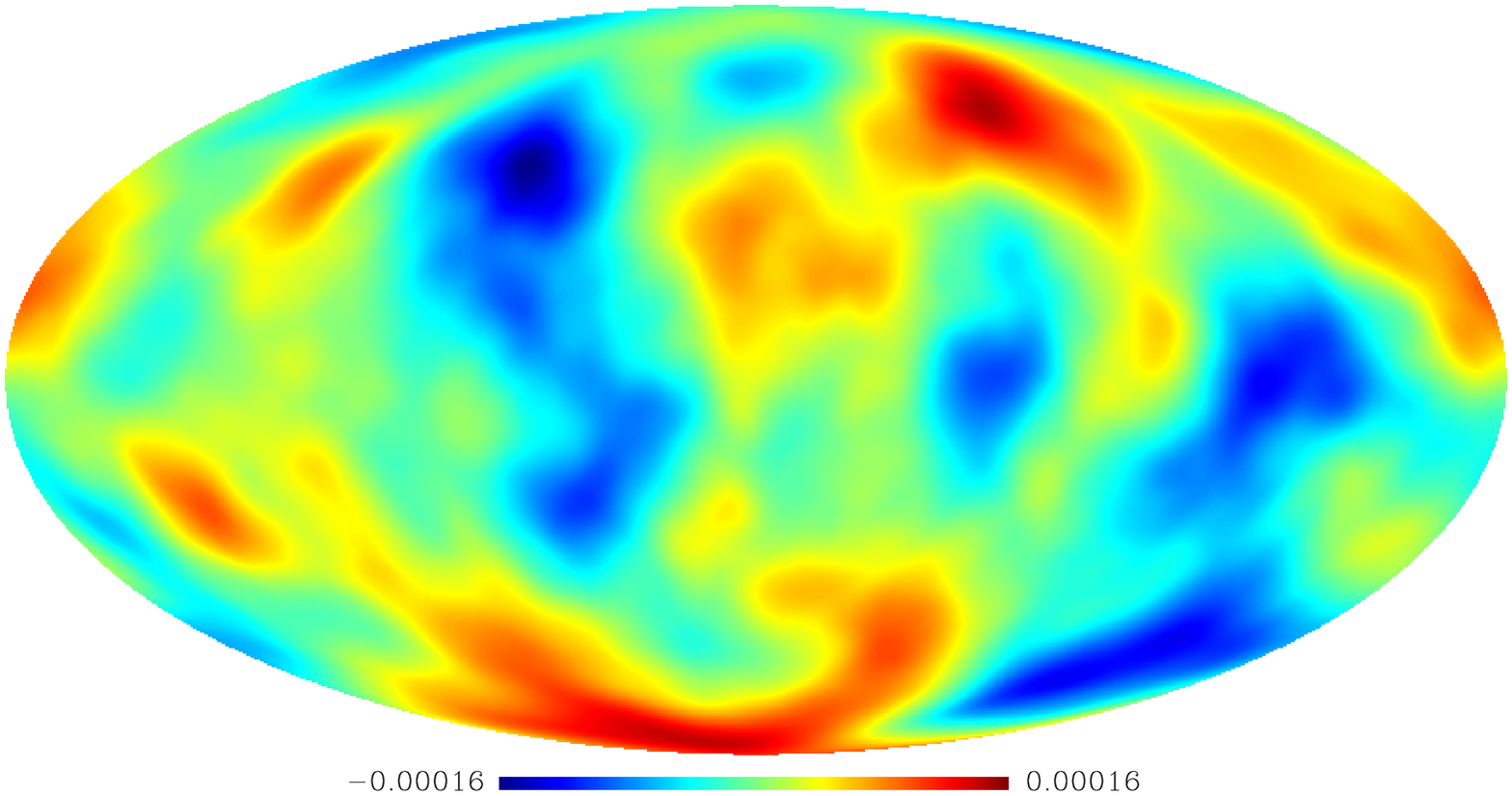} } %
{\includegraphics[width=8.8cm,height=6.5cm]{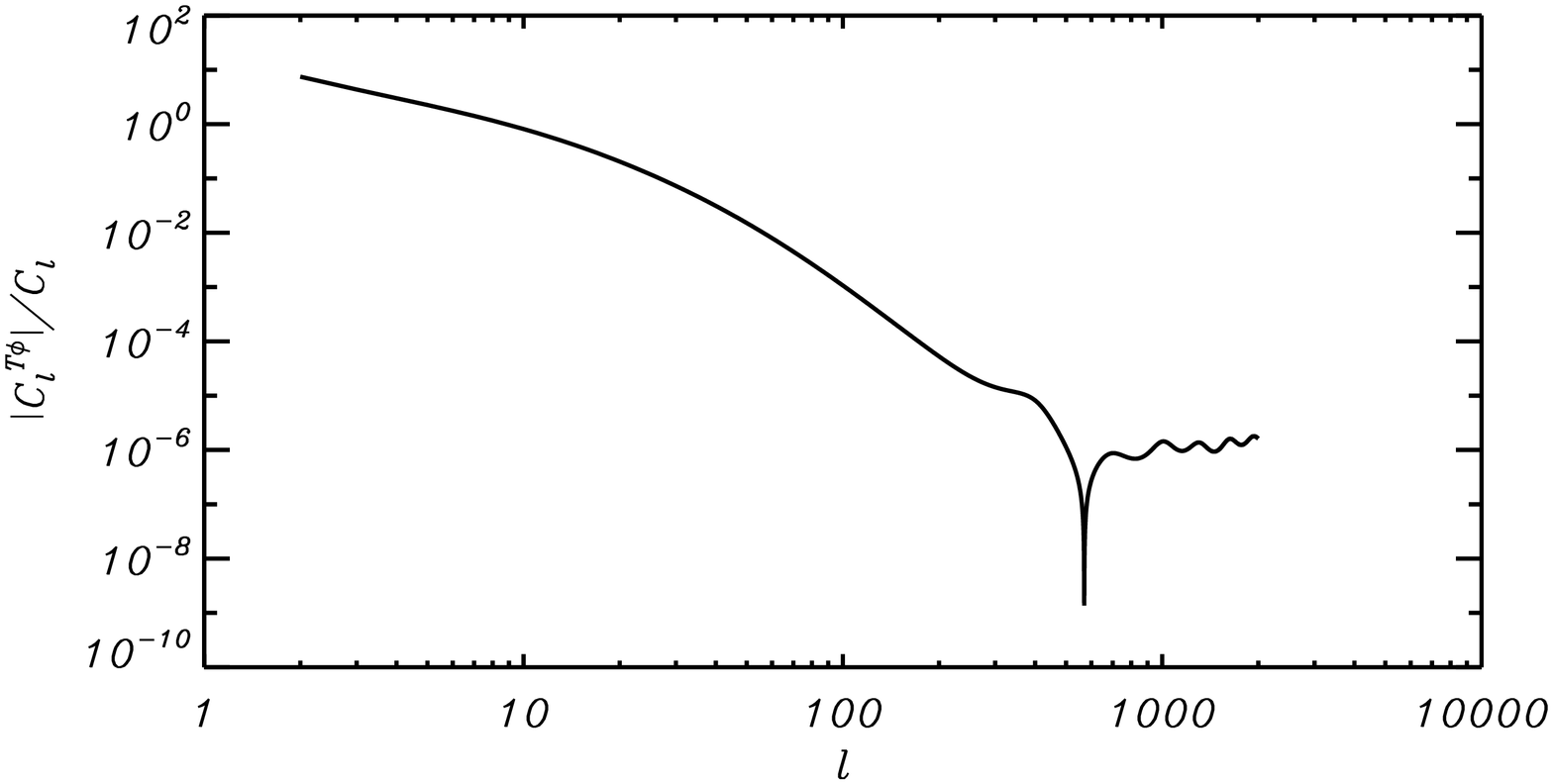} }
\caption{ {\it The Large scale contribution to the non-Gaussian L-ISW signal.} Upper panel: the map $Q(\hat{\bf{n}})\equiv  \sum_{\ell m} C^{T \phi}_\ell (C^{-1} a)_{\ell m} Y_{\ell m}(\hat{\bf{n}})$ contains the L-ISW coefficients $C^{T \phi}_\ell $ and enters the L-ISW estimator Eq. \ref{eq:lens-RS}. 
The $\ell$-filter $\frac{C_\ell^{T \phi}}{C_\ell}$ acts as a filter which suppresses the small scales (lower panel).}
\label{fig:Q}
\end{figure}

\begin{figure}
\centering
{\includegraphics[width=6.5cm,height=8.8cm,angle=90]{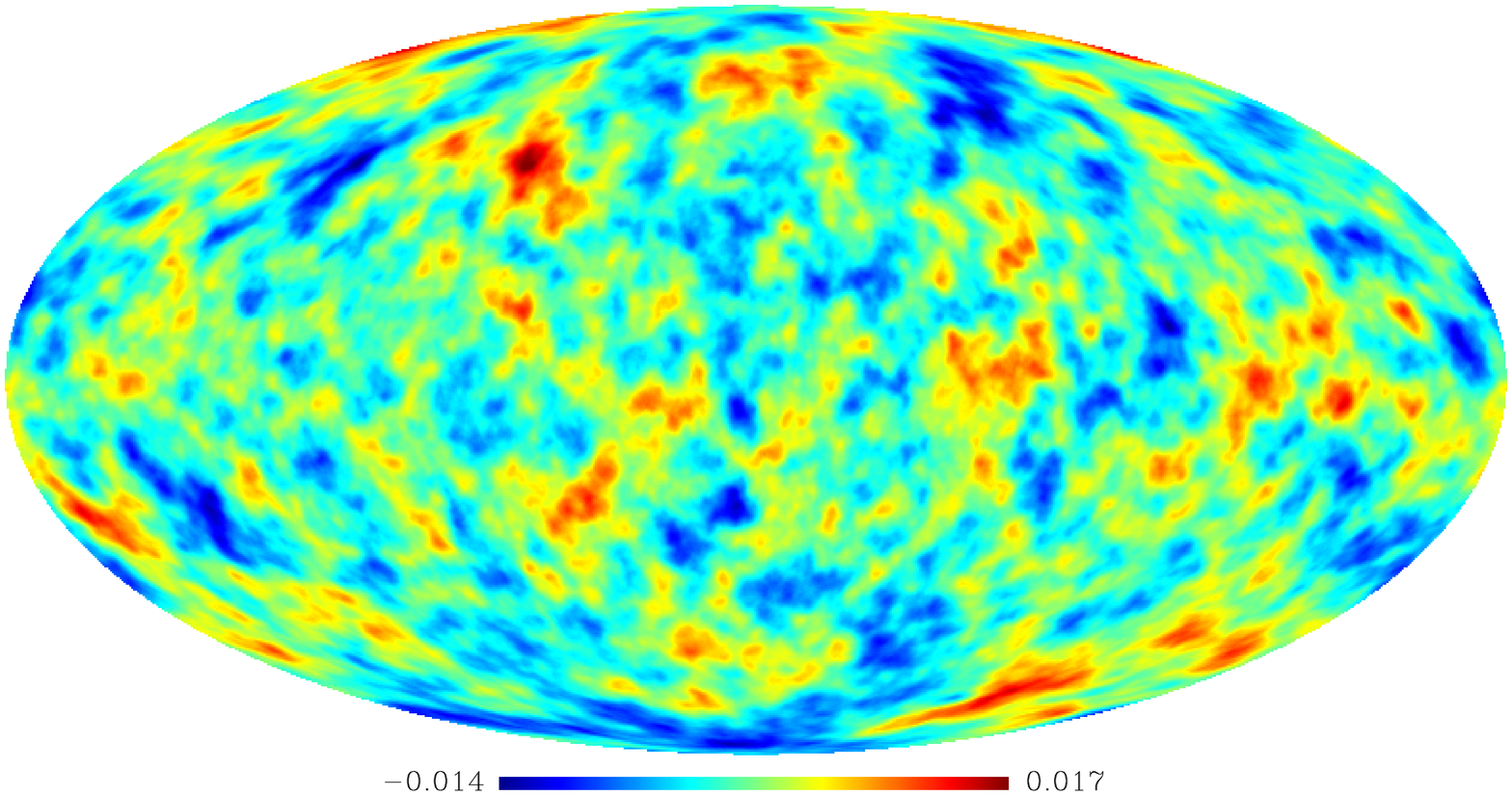} } \\
{\includegraphics[width=8.8cm,height=6.5cm]{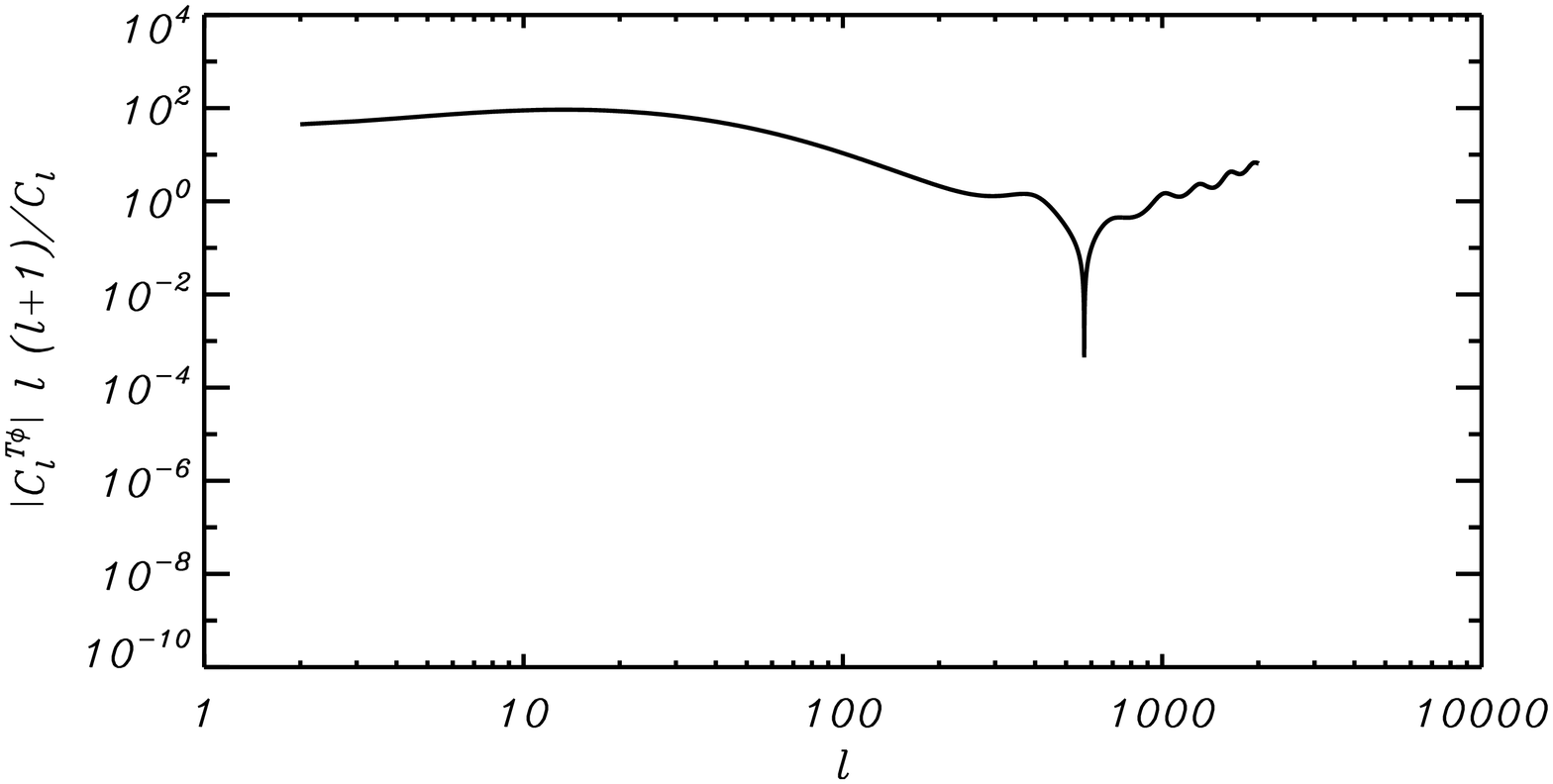}}
\caption{ {\it The intermediate scale contribution to the non-Gaussian L-ISW signal.} Same as figure \ref{fig:Q} but for the map $\delta^2 Q = -\sum_{\ell}\ell(\ell+1) C_\ell^{T \phi} (C^{-1} a)_{\ell m}Y_{\ell m}(\hat{\bf{n}})$ (upper panel) and its corresponding filter $-\ell(\ell+1) \frac{C^{T \phi}_\ell}{C_\ell}$.
The factor $\ell(\ell+1)$ dominates at high $\ell$ defining more small scale features with respect to the previous Q map. 
}
\label{fig:d2Q}
\end{figure}

Assuming that the only relevant non-Gaussian contribution is coming from the L-ISW term, which is the case if the local primordial non-Gaussianity is small and foregrounds and point sources have been correctly removed and masked, $S_{L-ISW}$ is given by the data as
\begin{eqnarray}
\nonumber
 S_{L-ISW} &\equiv &
\frac16\sum_{{all}~lm}
{\cal G}_{l_1l_2l_3}^{m_1m_2m_3}b_{l_1l_2l_3}^{L-ISW}
\Big[
(C^{-1}a)_{\ell_1m_1}(C^{-1}a)_{\ell_2m_2}(C^{-1}a)_{\ell_3m_3}\\
&-& 3(C^{-1})_{\ell_1m_1,\ell_2m_2}(C^{-1}a)_{\ell_3m_3}
\Big],
\label{eq:LISWest}
\end{eqnarray}
By factorizing the $\ell_i$ dependence, this becomes
\begin{eqnarray}
\nonumber
 S_{L-ISW}
&=& \frac12
\int d^2\hat{\bf{n}}
\left\{P(\hat{\bf{n}})[\delta^2 E](\hat{\bf{n}})Q(\hat{\bf{n}})\right.\nonumber\\
& &-[\delta^2 P](\hat{\bf{n}})E(\hat{\bf{n}})Q(\hat{\bf{n}})
-P(\hat{\bf{n}})E(\hat{\bf{n}})[\delta^2 Q](\hat{\bf{n}})
\left. \right\} \nonumber\\
&+&S_{lin}^{ISW} , %
\label{eq:lens-RS}
\end{eqnarray}
 where the maps $P(\hat{\bf{n}})$, $E(\hat{\bf{n}})$, $Q(\hat{\bf{n}})$ etc. are the same as defined in Eqs. \ref{eq:maps} and, in the case of a realistic experiment, they are convolved with the experimental window function $w_\ell$ so that, for example, $P(\hat{\bf{n}}) \equiv\sum_{\ell m}  w_\ell a_{\ell m}Y_{\ell m}(\hat{\bf{n}})$.

In Eq. \ref{eq:lens-RS}, the first two lines refer to the cubic part of the estimator, while $S_{lin}^{L-ISW}$ is the linear part which corrects for anisotropies and must be included in the case rotational invariance is not preserved. 
Details on the analytic expression of the L-ISW linear term and on its numerical implementation are given, respectively, in the next subsection \ref{sub:linear} and in Sec. \ref{sec:results}.

\subsection{The linear term}\label{sub:linear}


The linear term of the estimator is given by
\bea
S_{lin}^{L-ISW}&=& -\frac{1}{2} \int d^2 \hat{n}   \sum_{{all}~\ell m} b_{\ell_1 \ell_2 \ell_3}^{L-ISW} \\
&&(C^{-1})_{\ell_1m_1, \ell_2m_2}(C^{-1}a)_{\ell_3m_3} Y_{\ell_1 }^{m_1}({\bf{\hat{n}}}) Y_{\ell_2 }^{m_2}({\bf{\hat{n}}})  Y_{\ell_3 }^{m_3}({\bf{\hat{n}}}).\nonumber
\eea

By using the explicit form of $b_{\ell_1 \ell_2 \ell_3}^{L-ISW}$ and by factorizing the $\ell$-dependence one obtains 

\begin{eqnarray}
S_{lin}^{L-ISW} &=&  -\frac{1}{2} \int d^2 \hat{n}  \Big\{ Q( \hat{n}) \Big [ \langle P( \hat{n}) \delta^2E( \hat{n}) \rangle_{MC}  -  \langle E( \hat{n}) \delta^2P( \hat{n}) \rangle_{MC}  \Big ] \nonumber \\
                                                            &-& \delta^2Q( \hat{n}) \langle P( \hat{n}) E( \hat{n})) \rangle_{MC}  \\
                                                            &-& E( \hat{n})  \Big [ \langle Q( \hat{n}) \delta^2P( \hat{n}) \rangle_{MC}  -  \langle P( \hat{n}) \delta^2Q( \hat{n}) \rangle_{MC}  \Big ] 		\nonumber\\
                                                           &+& \delta^2E( \hat{n}) \langle P( \hat{n}) Q( \hat{n}) \rangle_{MC}  
                                                            -  \delta^2P( \hat{n}) \langle E( \hat{n}) Q( \hat{n}) \rangle_{MC} \nonumber \\
                                             &+& P( \hat{n})  \Big [ \langle Q( \hat{n}) \delta^2E( \hat{n}) \rangle_{MC}  -  \langle E( \hat{n}) \delta^2Q( \hat{n}) \rangle_{MC} \Big ]  \Big\}, \nonumber
  \label{eq:linear}
  \end{eqnarray}                                                         
where $\langle \rangle_{MC}$ indicates the Monte Carlo (MC) averages and the different maps are defined in Eq. \ref{eq:maps} and they are convolved with the experimental window function $w_\ell$, so that $P(\hat{\bf{n}}) \equiv\sum_{\ell m}  w_\ell a_{\ell m}Y_{\ell m}(\hat{\bf{n}})$, etc.

\begin{figure}
\includegraphics[width=8.8cm,height=6.5cm]{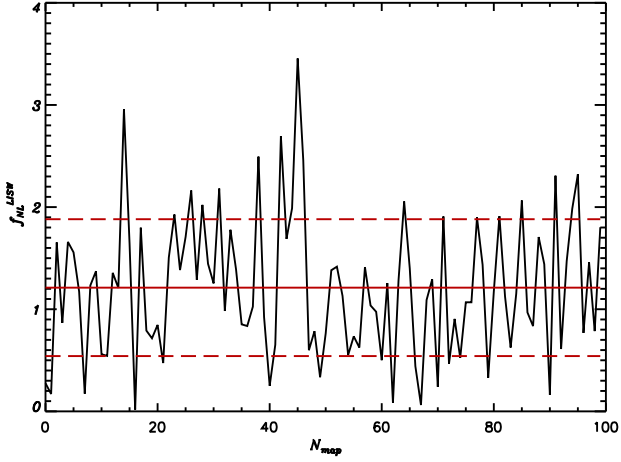} 
\caption{The plot shows the $f_{NL}^{L-ISW}$ values for 100 simulated non-Gaussian maps obtained with the covariance method of Sec. \ref{sec:COV}. The lensing part has been computed with the FLINTS code \citep{Lavaux:2010ja}. The straight line refers to the averaged $f_{NL}^{L-ISW}$ from these simulations, while the dashed line to the averaged 1-$\sigma$ error. Here $\ell_{max}=1000$. 
 }
 \label{Fig:fnlISW_cov1000}
\end{figure}

\subsection{Wiener filtered maps}\label{WFmaps}

The optimal bispectrum estimator as described in Eqs. (\ref{eq:LISWest}, \ref{eq:lens-RS})
involves products of inverse variance filtered maps, $C^{-1} a = (S +
N)^{-1} a$, where $S$ and $N$ are the signal and the noise covariance
matrix, respectively. A brute force calculation of such an expression
is impractical for modern high-resolution experiments as it involves
the inversion of two matrices that are too large to be stored and
processed as dense systems. In case the noise covariance can be
described in terms of a simple power spectrum in spherical harmonic
space, the calculation simplifies significantly.
However, this approach is no
longer exact for experiments with anisotropic noise distribution or
reduced sky coverage, leading to an increase in the error bars of the
estimates.

Here, we use Wiener filtering as a basis for the exact evaluation of
terms involving $C^{-1} a$. We apply the iterative scheme of
\citep{Elsner:2012fe} to calculate the Wiener filter $a^{\mathrm{WF}}
\equiv S (S + N)^{-1} a$, the maximum a posteriori solution in case
signal and noise are Gaussian random fields. After $a^{\mathrm{WF}}$
has been successfully computed, we finally obtain the inverse
variance filtered map by normalizing the spherical harmonic
coefficients of the Wiener filter solution by the CMB power spectrum
multiplied with the beam window function, $C^{-1} a_{\ell m} = (C_\ell^{th}
b_\ell^2)^{-1} a^{\mathrm{WF}}_{\ell m}$.
%

\section{Results}\label{sec:results}

In this section we present the results regarding the numerical implementation %
of the optimal estimator and of the methods presented in sec. \ref{sec:COV} and \ref{sec:NGmethod} to build the CMB maps containing the L-ISW bispectrum. 
In particular, we processed the simulated L-ISW maps through the estimator pipeline to get the amplitude parameter $f_{NL}^{L-ISW}$ of Eq. \ref{eq:est}.
We consider two main settings: 
\begin{itemize}
\item a full sky cosmic variance limited CMB experiment %
up to a maximum multipole $\ell_{max} \simeq 1000$ and 
\item a more realistic experimental setting which consists of a one channel CMB experiment with a Gaussian beam with a FWHM $\theta_b=7'$, a galactic mask leaving $\simeq 80\%$ of the sky and anisotropic uncorrelated noise.  
These settings are visualized in Figs. \ref{Fig:expsetting1}, \ref{Fig:expsetting2} and details are given in Sec. \ref{app:expsetting}. 
\end{itemize}
 All runs have been performed at full resolution $N_{side}=2048$ (which corresponds to a map pixel number of $5.033 \cdot 10^7$). The maps in Eqs. (\ref{eq:maps}) are calculated by using the Healpix package \citep{Gorski:2004by}. 
 The theoretical power spectrum of the temperature-only primary CMB coefficients $C_\ell$ has been generated with the CAMB code for a fiducial $\Lambda$CDM cosmological model with parameters corresponding to WMAP7 cosmological parameters \citep{Komatsu:2010fb}.
For illustrative purpose, the plots of the maps and of the correspondent $\ell$-filters containing the L-ISW cross correlation coefficients $C_\ell^{T \phi}$ 
are shown in Figs. \ref{fig:Q} and \ref{fig:d2Q}. 


\begin{figure}
\includegraphics[width=8.8cm,height=6.5cm]{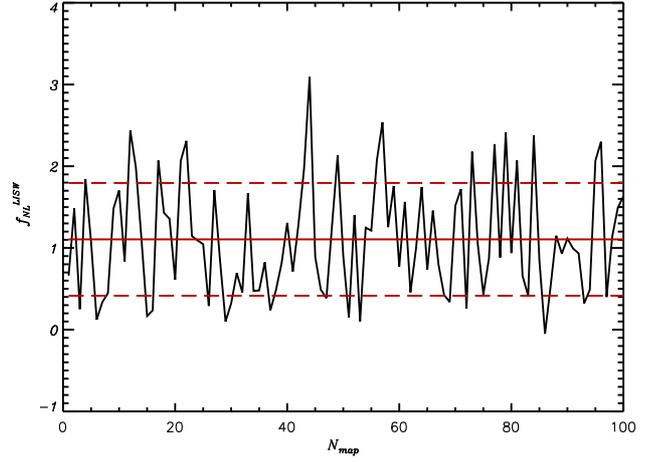}
\caption{Same as Fig. \ref{Fig:fnlISW_cov1000} but for 100 simulations built with the separable mode expansion method  (Eq. (\ref{eq:almNG})).
}
\label{fig:fnl-almNG}
\end{figure}

We built a set of 100 CMB simulations for each of the two methods described in sections \ref{sec:NGmethod} and \ref{sec:COV} for a cosmic variance limited CMB experiment with full sky coverage.
For the covariance method, we used the FLINTS code \citep{Lavaux:2010ja} to generate the lensing coefficients $a_{\ell m}^L$ and the lensing potential coefficients $\phi_{\ell m}$ needed to build the non-Gaussian $a^{L-ISW}_{\ell m}$ as described in Sec. \ref{sec:COV}.
In both cases, we analyzed the L-ISW simulated CMB maps with the L-ISW estimator up to $\ell_{max}=1000$. 
According to the definition of $f_{NL}^{L-ISW}$, the expected value is 1 with 1-$\sigma$ error predicted from theory for  $\ell_{max}=1000$ of $\simeq 0.64$.  
In the case of the separable expansions mode method, the simulations give a mean  $f_{NL}^{L-ISW} =1.1$  with averaged 1-$\sigma$ error $\simeq 0.69$.  
With the simulations built with the covariance method, we obtain a mean $f_{NL}^{L-ISW}  = 1.21$ with averaged 1-$\sigma$ error of 0.67.
 The results are summarized in Fig. \ref{fig:fnl-almNG} and Fig. \ref{Fig:fnlISW_cov1000}, respectively. 
 These estimates are compatible with the theoretical predictions.
The error bars are slightly suboptimal because of numerical noise and the fact that we are assuming a diagonal covariance matrix so that $(C^{-1} a)_{\ell m}=a_{\ell m}/C_\ell$. 


\begin{figure}
\includegraphics[width=8.8cm,height=6.5cm]{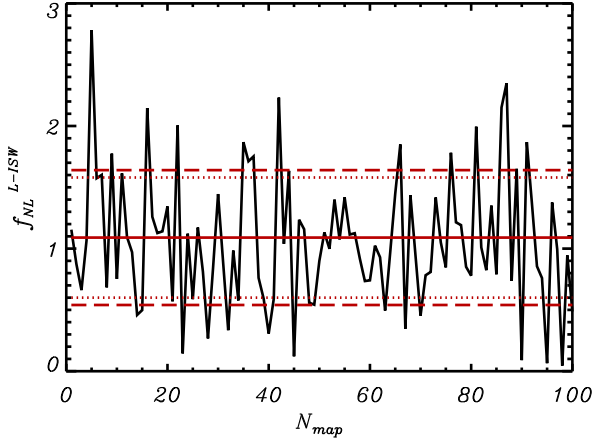} %
\caption{The same as Fig.\ref{fig:fnl-almNG} but for a more realistic CMB experiment with a 7' FWHM Gaussian beam, anisotropic noise and 20\% galactic mask.  Here $\ell_{max}=1500$. The dashed lines are the 1-$\sigma$ averaged error bars from simulations while the dotted lines are the expected Fisher errors. 
}
\label{fig:fnl-almNGnoisebeam}
\end{figure} 

\begin{figure}[!t]

\includegraphics[width=8.8cm,height=6.5cm]{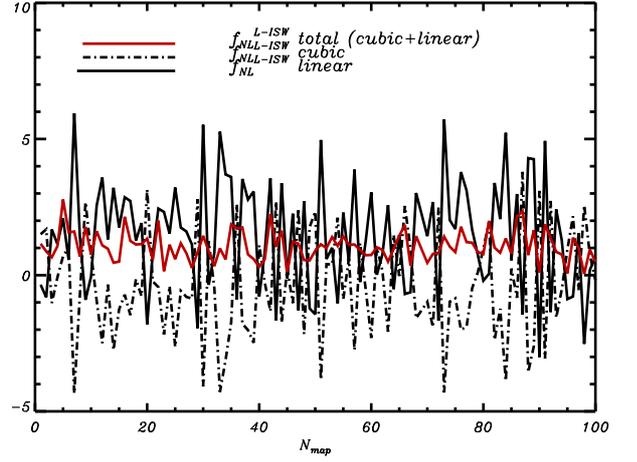}%
\caption{The linear term of the estimator reduces the error bars in the case of anisotropic data. The plot shows the linear (solid black line) and the cubic (dot-dashed black line) contributions to the total (red line) $f_{NL}^{L-ISW}\equiv  (f_{NL}^{L-ISW})_{cubic}+ (f_{NL}^{L-ISW})_{linear}$ in the case of a CMB experiment with anisotropic noise and 22\% galactic mask. 
}
 \label{Fig:lineartest}
\end{figure}

In order to test the estimator on a more realistic case, we built a set of 100 simulations with the separable mode expansion method considering a realistic experimental setting. This consists of a CMB one channel experiment with a Gaussian beam FWHM $\theta_b=7'$, a galactic mask with $f_{sky} = 0.78$ and anisotropic noise, as previously described.
In this case, we run the estimator up to $\ell_{max}=1500$. 
The expected theoretical 1-$\sigma$ error on $f_{NL}^{L-ISW}$ for this experimental setting and up to $\ell_{max}=1500$ is $\simeq 0.49$. This estimate accounts for a $\simeq 10\%$ percent increase in the error bar due to the fact that the lensing is intrinsically non-Gaussian and it gives an extra contribution to the variance, as shown in \citep{Lewis:2011fk}. We get a mean  $f_{NL}^{ISW} =1.09$ with averaged 1-$\sigma$ error $\simeq 0.55$. In this case we computed both the cubic and linear part of the estimator. 
In particular, the linear term has been tested with a set of 100 Monte Carlo (MC) averages generated for each map product in equation (Eq. \ref{eq:linear}).
In the presence of anisotropic noise and a sky cut, the linear part of the estimator is necessary to recover the expected estimation of $f_{NL}
^{L-ISW}$ and error bars. The linear contribution to $f_{NL}^{L-ISW}$ is strongly anti-correlated with the cubic part.
This behavior is summarized in figure Fig. \ref{Fig:lineartest}. 
In the plot are shown the linear and the cubic contributions to the total amplitude $f_{NL}^{L-ISW}\equiv  (f_{NL}^{L-ISW})_{cubic}+ (f_{NL}^{L-ISW})_{linear}$. We also checked that with 100 MC averages the linear term converges and it is stable: for this specific experimental setting we find that the results do not improve when increasing the MC averages to 200.

Finally, in order to test optimality, we Wiener filtered the 100 L-ISW simulations and we processed them through the L-ISW estimator pipeline. 
 The maps has been produced following \citep{Elsner:2012fe}, as described in section Sec. \ref{WFmaps}. %
 We use as inputs the same experimental settings as described previously.
The linear term has been computed with 100 Wiener filtered MC simulations.
 We found that the improvement with respect to the non Wiener filtered simulations is small ($< 10\%$) in the case of our particular settings. However, this does not exclude that the Wiener filtering may have a more noticeable  impact for a more realistic experimental setting and noise covariance.

\section{$f_{NL}$ error estimation}\label{sec:err}

This section summarizes the results regarding the impact of the L-ISW signal on the error estimation of  $f_{NL}$ from the local type non-Gaussianity.
If the only contribution to $f_{NL}$ were from the primary local type non-Gaussianity the error on this parameter would be simply given by
\be
\sigma^P=\sqrt{\frac{1}{F^P}},
\ee
i.e. the inverse of the Fisher matrix of the local type non-gaussian contribution
\be
\label{eq:fishP}
F^{P}=  f_{sky}\sum_{2 \leqslant \ell_1 \leqslant \ell_2 \leqslant \ell_3} \frac{ B^{P}_{\ell_1 \ell_2 \ell_3} \, B^{P}_{\ell_1 \ell_2 \ell_3} }{ \Delta_{\ell_1 \ell_2 \ell_3} C_{\ell_1}  C_{\ell_2}   C_{\ell_3} },
\ee
where $f_{sky}$ refers to the observed sky fraction. The noise, $N_\ell$, and the beam, $b_\ell$, can be accounted for so that $C_\ell=N_\ell+C_\ell^{th}b^2_\ell$. In this case the bispectrum is also convolved with the beam transfer function $b_\ell$: $B^{P}_{\ell_1 \ell_2 \ell_3} \propto b^{P}_{\ell_1 \ell_2 \ell_3}b_{\ell_1}b_{\ell_2}b_{\ell_3} $.  

However, the L-ISW can be a serious contaminant of the local primary signal \citep{Mangilli:2009dr,Hanson:2009kg}, so that it is important to quantify the effect on the expected $f_{NL}$ error as well.
If the L-ISW signal is present, the error matrix will be given by a non-diagonal Fisher matrix of the form
\be
F_{ij}= \left(
  \begin{array}{cc}
  F^P&F^{cross}\\
  F^{cross}&F^{L-ISW},
  \end{array}
  \right)
\ee
 where
 \be
F^{cross}=f_{sky}\sum_{2 \leqslant \ell_1 \leqslant \ell_2 \leqslant \ell_3} \frac{ B^{L-ISW}_{\ell_1 \ell_2 \ell_3} \, B^{P}_{\ell_1 \ell_2 \ell_3} }{ \Delta_{\ell_1 \ell_2 \ell_3} C_{\ell_1}  C_{\ell_2}   C_{\ell_3} },
\ee
is the cross correlation term and $F^{L-ISW}$ is the the Fisher term of the L-ISW signal of Eq. \ref{eq:FisherLISW}.
The expected error on the local $f_{NL}$ will be then:
\be
\sigma^P_{cross}=\sqrt{(F^{-1})_{11}},
\ee
i.e. the inverse of the full Fisher matrix containing the cross correlation between the primary local non-Gaussianity and the L-ISW signal.
The difference between the error estimation on $f_{NL}$-primary with and without the L-ISW contribution is:
\be
\Delta \sigma^P=\sigma^P_{cross} - \sigma^P.
\ee
To quantify the level of correlation between the two signals one can define the correlation coefficient as %
\be
r=\frac{F^{cross}}{\sqrt{F^{P}} \sqrt{F^{L-RS}}}.
\ee
We find that the effect of the local non-Gaussianity on the L-ISW is negligible. Therefore, the $1-\sigma$ error of the L-ISW amplitude parameter $f_{NL}^{L-ISW}$ is given by $\sigma^{L-ISW}=\sqrt{(F^{L-ISW})^{-1}}$.

For a realistic CMB experiment as described in Sec. \ref{sec:results} and appendix \ref{app:expsetting}, we find that the correlation between the two signals is $r=0.20$ at $\ell_{max}=1500$ and $r=0.27$  at $\ell_{max}=2000$ 
and that the expected detection significance of the L-ISW signal ($1/\sigma^{L-ISW}$) is, respectively, at $\simeq 2$ and $\simeq 3$ $\sigma$.   
The effect on the $f_{NL}$ local error due to the contamination is in the range between $\simeq 3\%$ to $\simeq 5$ \% for $\ell_{max}$ from 1000 to 2000, depending on how much the two signal are correlated.
  For $\ell_{max}=2000$, if the L-ISW signal is not accounted correctly,  the $f_{NL}$ error bars are overestimated by $\simeq 4\%$.

\section{Discussion and conclusions}\label{sec:conclusions} 
We have presented the formalism and the numerical implementation to build  the optimal KSW estimator for the Lensing-ISW bispectrum. Moreover, we have tested the estimator on simulated CMB maps containing the Lensing-Integrated Sachs Wolfe (L-ISW) non-Gaussian signal and on the Wiener filtered simulations in order to test optimality.
As regarding simulations, we have implemented and tested two methods: a non-perturbative approach to simulate CMB sky maps with the L-ISW signal which is based on the FLINTS lensing code \citep{Lavaux:2010ja} and the perturbative separable mode expansion method calculated for this specific signal. 
 We provide the analytical expression and the numerical implementation of the linear term of the estimator for this specific kind of bispectrum. 
For a realistic CMB experimental setting accounting for anisotropic noise and masked sky, the linear term gives a relevant contribution which 
is highly anti-correlated with the cubic part 
and it is necessary to recover the signal and optimal error bars.
 In order to achieve optimality, we also tested the estimator on the Wiener filtered L-ISW simulated CMB maps. In this case we recovered the signal with error bars which saturate the theoretical Cramer-Rao bound, with a small  improvement of $< 10\%$ with respect to the non Wiener filtered simulations. 
Finally,  we estimate that, if not correctly accounted for, the L-ISW effect has also an impact on $f_{NL}^{local}$ error bars leading to a bias and an overestimation of $\simeq 4\%$, in agreement with \citep{Lewis:2011fk}. 
 Thus a joint analysis of non-Gaussian shapes and/or L-ISW template subtraction will be needed in order to recover unbiased minimum variance results of the local type primordial non-Gaussian signal. 

It is important to note that the KSW bispectrum approach to the estimation of the L-ISW is complementary to the lensing reconstruction estimator of \citep{Lewis:2011fk}.
 In principle, the KSW estimator can offer advantages with respect to other methods 
 because the bispectrum has a unique shape and it 
 has been shown to be robust to foreground contamination \citep{Yadav:2010fz} 
 so it can be measured by using a larger sky fraction. In addition, inclusion of the L-ISW in the framework of bispectrum analysis gives an unified approach to testing for primordial non-Gaussianity.
 The tools presented in this paper enable the optimal analysis of this important signal from future CMB data.
 

\begin{acknowledgements} 
This work was supported in part by
NSF grants AST 07-08849 and AST 09-08902, and  by NASA/JPL subcontract 1413479; and through Ben Wandelt's ANR Chaire d'excellence ANR-10-CEXC-004-01.
AM acknowledges  Guilhem Lavaux for the FLINTS lensing simulations, Licia Verde for useful comments and discussion and the University of Illinois for the use of the curvaton computers.
 \end{acknowledgements}



\bibliographystyle{aa}

\bibliography{lisw_aa}

\begin{appendix}

\section{The simulated L-ISW CMB bispectrum from the covariance method}\label{app:ISWcov}
 This appendix refers to the covariance method used to build the L-ISW simulated maps and described in Sec. \ref{sec:COV}.
It is straightforward o check that the coefficients $ a^{L-ISW}_{\ell m}=z_{\ell m} + \Delta a^L_{\ell m} $ give the wanted bispectrum by calculating $ \langle  (a^{ISW}_{\ell m})^3\rangle = \langle(z_{\ell m} + \Delta a^L_{\ell m})^3 \rangle$.
The lensing coefficients $a_{\ell m}^{ L}$ can be expressed analytically, at first order in the lensing expansion, as 
\bea%
a_{\ell n}^{ L} &=&  a_{\ell m}^{P}+\sum_{\ell'\ell''m' m''}(-1)^{m+m'+m''}{\cal G}^{-m m' m''}_{\ell \ell' \ell''} \\
&&  \frac{\ell'(\ell'+1)-\ell(\ell+1)+\ell''(\ell''+1)}{2}a^{P*}_{m'\ell'}\phi^{* L}_{ \ell'' -m''}, \nonumber
\eea
 where $a_{\ell m}^{P}$ the primary and $\phi^{L}_{ \ell m}$ the harmonic coefficients of the lensing potential $\phi^L$. Since, according to the new variables definition $(z_{\ell m},t_{\ell m})$ of eq. \ref{eq:ISWalm},
  $t_{\ell m}=\phi^{L}_{ \ell m}$, the $a_{\ell m}^{ L}$ can be written as:
\be
a_{\ell m}^{L} \propto a_{\ell m}^{P}+f_\ell a_{\ell m}^{P *} t_{\ell m}.
\ee
Here, for simplifying the notation, $f_\ell= \sum_{\ell'\ell''m' m''}(-1)^{m+m'+m''}{\cal G}^{-m m' m''}_{\ell \ell' \ell''} \frac{\ell'(\ell'+1)-\ell(\ell+1)+\ell''(\ell''+1)}{2}$ so that $ \Delta a^L_{\ell m}=f_\ell a_{\ell m}^{P *} t_{\ell m}$ at first order.
  The explicit expression for  $\langle  (a^{L-ISW}_{\ell m})^3\rangle$ takes the form:
 \bea
 \langle  (a^{LISW}_{\ell m})^3\rangle &=& \langle(z_{\ell m} + \Delta a^L_{\ell m})^3 \rangle=\langle z_{\ell m}^3\\
 &+&z_{\ell m}( \Delta a^L_{\ell m})^2 +3 z_{\ell m}^2  \Delta a^L_{\ell m} 
        +( \Delta a^L_{\ell m})^3 +2 z_{\ell m}( \Delta a^L_{\ell m})^2 \rangle \nonumber
 \eea

From this the only non-zero term is:    %
 \bea
\langle 3 z_{\ell m}^2  \Delta a^L_{\ell m}\rangle&=&
3\langle f_\ell a_{\ell m}^{P*} t_{\ell m} C_\ell^{TT}\Big(x_{\ell m} x_{\ell' m'}(r_\ell^{T\phi})^2 + y_{\ell m}y_{\ell' m'}(1-(r_\ell^{T\phi})^2) \nonumber\\
&+&2 x_{\ell m}r_\ell^{T\phi} y_{\ell' m'}\sqrt{1-(r_\ell^{T\phi})^2}\Big) \rangle  
\eea
From this only survives:
\be
6\langle f_\ell a_{\ell m}^{P*} t_{\ell m} C_\ell^{TT} x_{\ell m}r_\ell^{T\phi} y_{\ell' m'}\sqrt{1-(r_\ell^{T\phi})^2} \rangle.
\ee
By using the definition of $r_\ell^{T \phi}$ in Eq. \ref{eq:r} and the approximation $r_\ell^{T \phi}<<1$  for which $1-(r_\ell^{T\phi})^2 \simeq 1$, since by construction: $ a_{\ell m}^{P}= y_{\ell m}\sqrt{C_\ell^{TT}}$, $t_{\ell m}=x_{\ell m}\sqrt{C_\ell^{\phi \phi}}$, $\langle x^2 \rangle=1$,  $\langle y^2 \rangle=1$ and $\langle xy \rangle=0$ we recover the expected signal: 

 \be
 \langle  (a^{LISW}_{\ell m})^3\rangle = 6 f_\ell C_\ell^P C_\ell^{\phi T}. %
\ee 
\end{appendix}

\begin{appendix}
\section{L-ISW cross correlation coefficients}\label{app:Cltphi}
The definition of the CMB lensing-ISW/RS cross correlation coefficients is \citep{Spergel:1999xn,Verde:2002mu,Giovi:2004te}:
\be
C^{T \phi}_\ell \equiv \langle \phi_{L \,\ell}^{*m} a_{\ell}^{
m}\rangle 
\simeq 2\int_0^{z_{ls}}\!\!\frac{r(z_{ls})-r(z)}{r(z_{ls})r(z)^3}.
\left[
\frac{\partial}{\partial z}P_{\phi}(k,z)
\right]_{k=\frac{\ell}{r(z)}}\!\!\!dz. %
\label{eq:cltphi}
\ee
where, $r(z)$ is the co-moving conformal distance and $P_{\phi}(k,z)$ is the gravitational potential power spectrum which accounts for both the linear and non-linear contributions.
The non-linear regime RS contribution to the signal is tiny, in agreement with \citep{Lewis:2012tc,Junk:2012qt} 
\footnote{We found that the amplitude on the non-linear effect estimated in \citet{Mangilli:2009dr} (as well as probably in \citet{Giovi:2004te})  compared to  \citet{Lewis:2012tc, Junk:2012qt} was due to an interpolation issue. Even a small numerical effect at the interpolation scale $k=\frac{\ell}{r(z)}$ can propagate through the line-of-sight integration to a relevant effect on the non-linear transition scale of the $C^{T \phi}_\ell$. Note however that the $C^{T \phi}_\ell$ coefficients are very sensitive to the cosmology parameters related to the late time evolution, $\Omega_\Lambda$, $w$, $\sigma_8$, and to the modeling of the non-linearities (e.g. \citet{Verde:2002mu}), so extra care must be taken when comparing results from different authors.}.
Considering both the linear ISW and the Rees Sciama effect improves the $f_{NL}^{L-ISW}$ variance and signal to noise by few percent ($\simeq 2 \%$) 
 with respect to the linear only case calculation. In this work we consider both contributions for completeness.  
As a template, for both the simulations and the estimator, we used the late ISW-lensing cross correlation coefficients of Eq. \ref{eq:cltphi}. This is a good approximation since this effect is the one which gives the main contribution. However, for a detailed description see \citep{Lewis:2012tc}. 
 
 \section{The experimental setting}\label{app:expsetting}
 
  \begin{figure}
{\includegraphics[width=8.8cm,height=6.5cm]{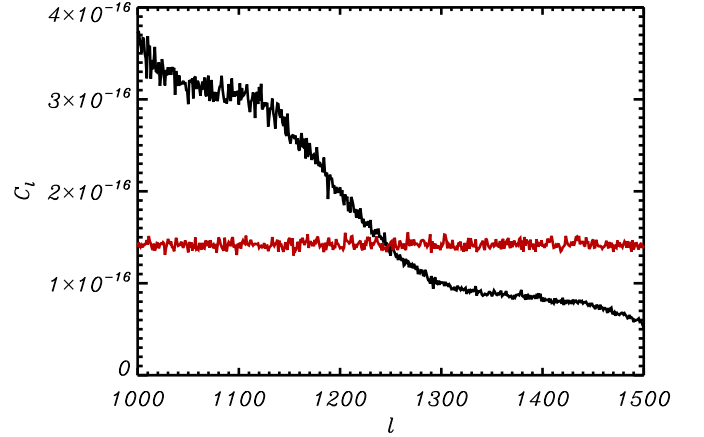}}\\ %
\centering
{\includegraphics[width=6.5cm,height=8.8cm,angle=90]{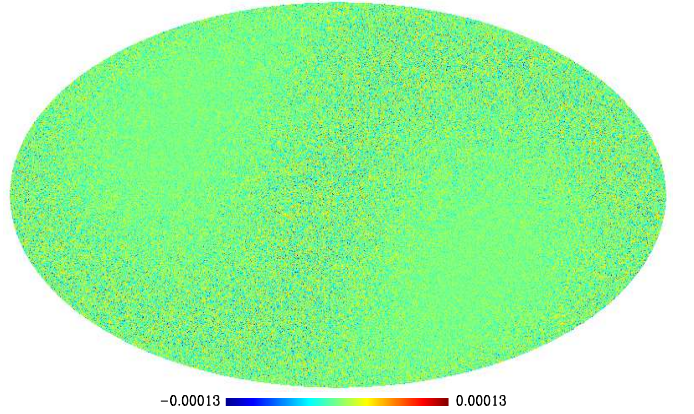}} %
\caption{{\it Noise simulations.} The figure shows an example of one simulated anisotropic noise map realization (bottom panel) and its correspondent noise power spectrum in red on the upper panel. The black line corresponds to the power spectrum  from the same L-ISW simulation. %
}
  \label{Fig:noiserealiz}
\end{figure}
 
   \begin{figure}[t!]
\centering
{\includegraphics[width=8.8cm,height=6.5cm]{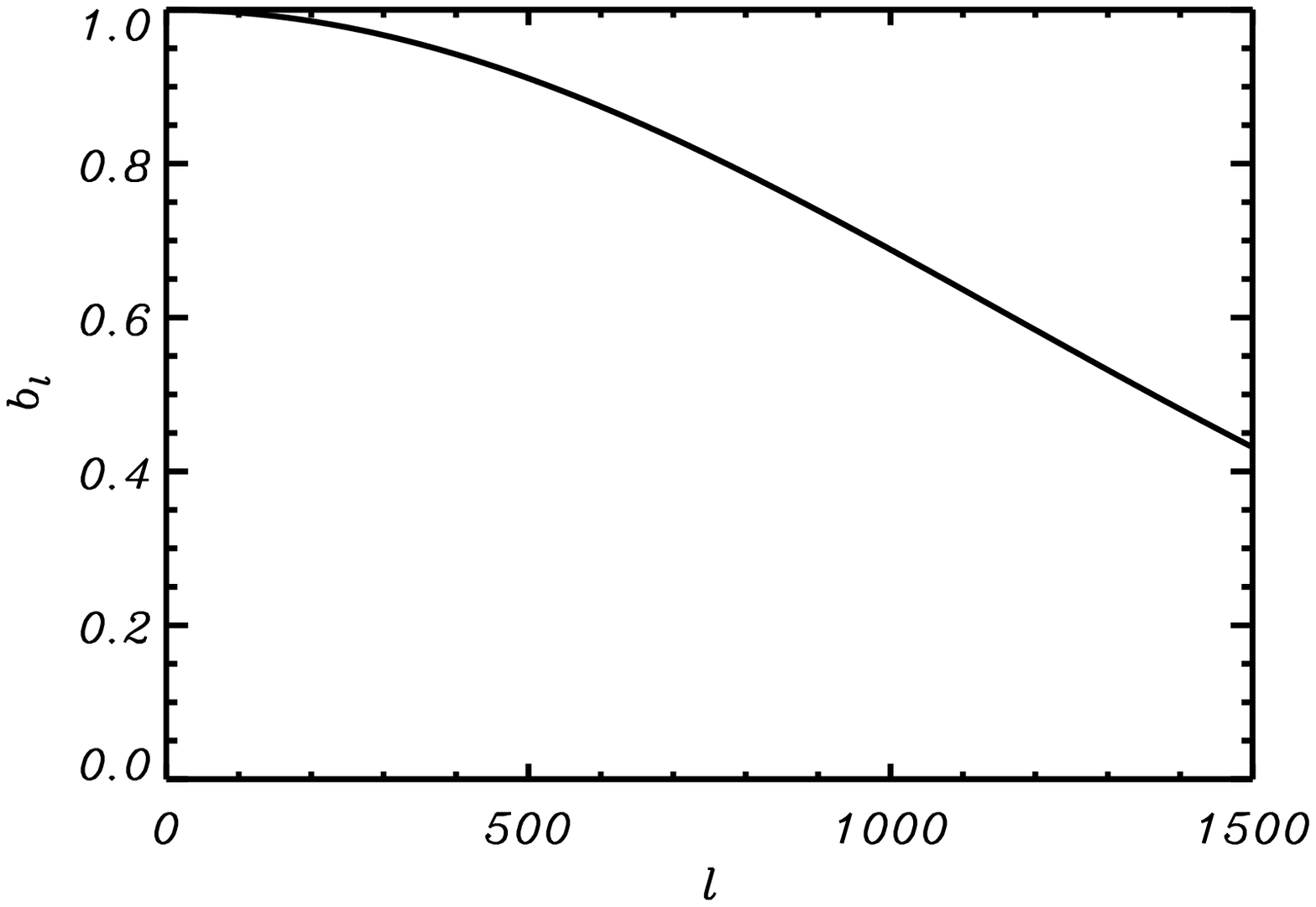}} \\
{\includegraphics[width=6.5cm,height=8.8cm,angle=90]{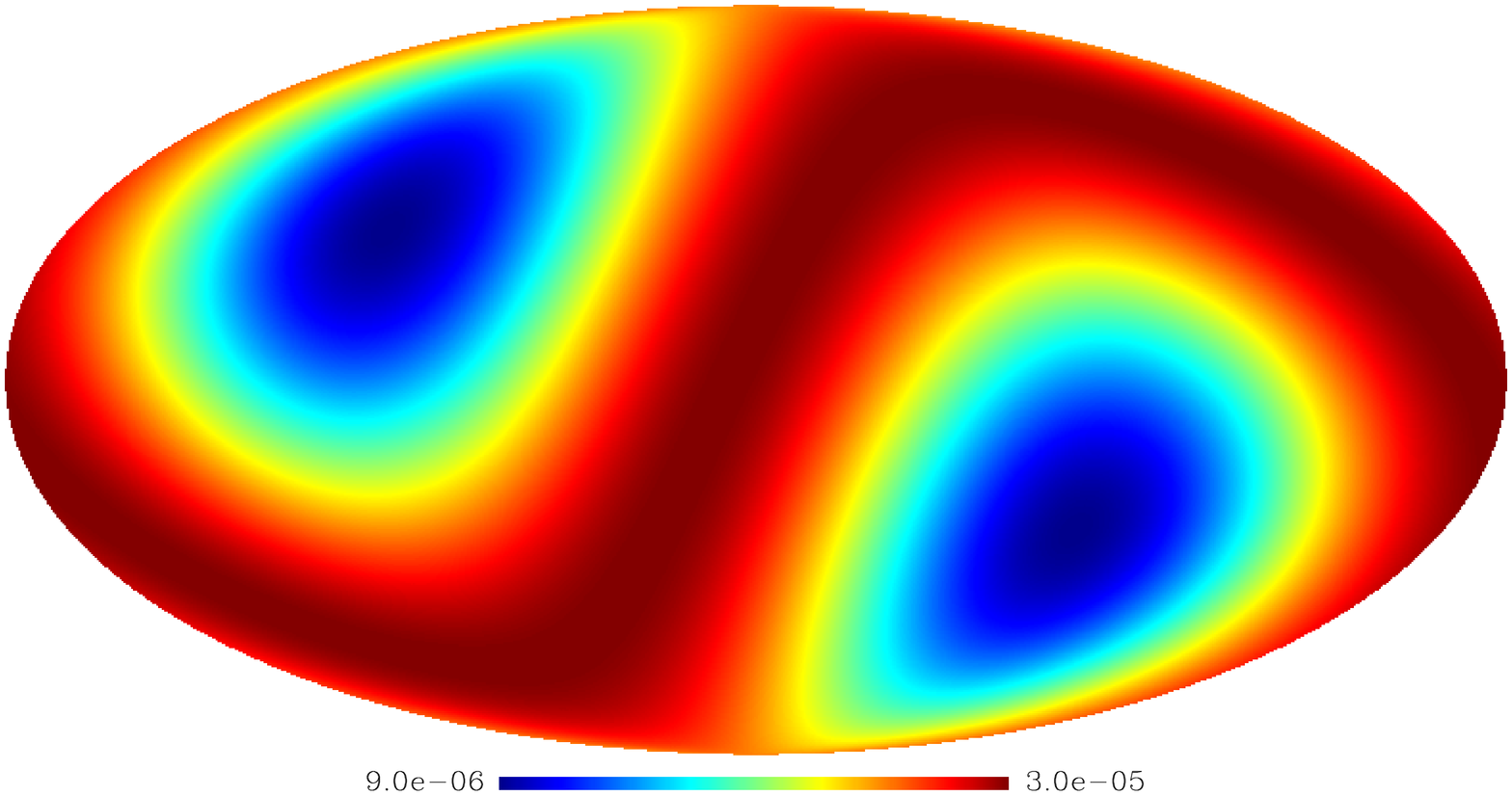}}
\caption{{\it Experimental setting: Beam window function and anisotropic noise map.} The Gaussian beam with a 7' FWHM is shown in the upper panel, while the bottom panel shows the dipole like anisotropic noise covariance matrix map.
}
 \label{Fig:expsetting2}
\end{figure}

\begin{figure}[t!]
\includegraphics[width=6.5cm,height=8.8cm,angle=90]{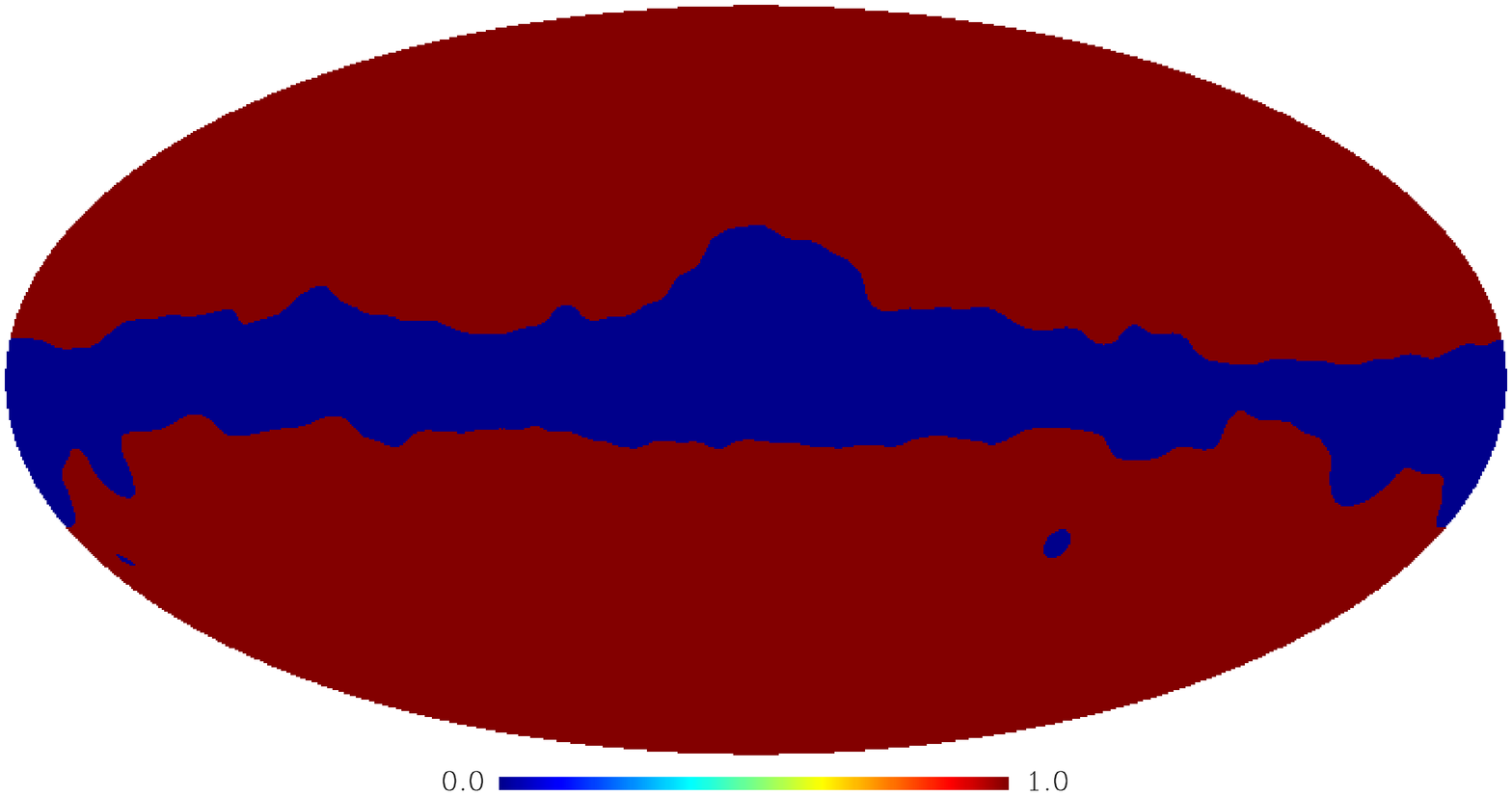}
\includegraphics[width=6.5cm,height=8.8cm,angle=90]{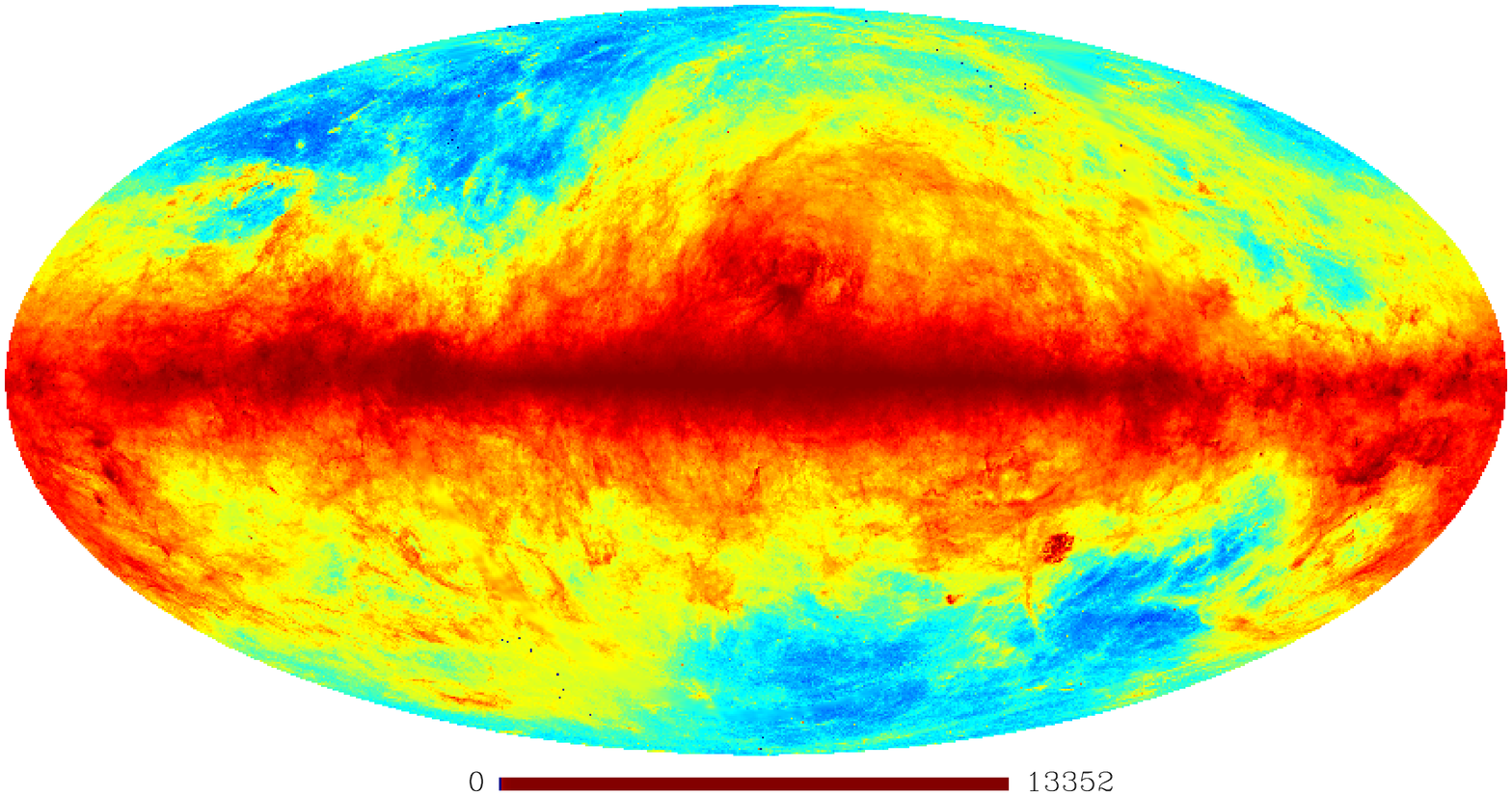}
\caption{{\it Mask.} Galactic mask cut with $f_{sky}=0.78$ (upper panel) obtained from thresholding the smoothed 100$\mu$m IRAS map (bottom panel). 
}
 \label{Fig:expsetting1}
\end{figure}

 Figs. \ref{Fig:expsetting2} and \ref{Fig:expsetting1} summarize the experimental settings used in the simulations. 
 These settings are inspired by a space-based experiment such as WMAP or Planck, with a variation of the noise with the ecliptic latitude.
 We consider a one channel CMB experiment with a Gaussian beam with a FWHM $\theta_b=7'$, a galactic mask leaving $\simeq 80\%$ of the sky and anisotropic uncorrelated noise.  
 In particular we obtain a galactic type mask from the IRAS\footnote{http://www.cita.utoronto.ca/~mamd/IRIS/IrisOverview.html 
} 100$\mu$m map smoothed at 5 angular degrees resolution and with a threshold of 12MJy/sr. We consider a dipole type anisotropic noise covariance matrix, which accounts for the anisotropies owing to, e.g., the scanning strategy.  An example of an anisotropic noise realization and the correspondent power spectrum is given, respectively, in the bottom and in the upper panels of  Fig. \ref{Fig:noiserealiz}. The noise start dominating from $\ell \simeq 1300$.

 \end	{appendix}
 
 
\end{document}